\documentclass[aps,prl,twocolumn,showpacs,showkeys,superscriptaddress]{revtex4}
\usepackage{graphicx,dcolumn,bm,amsmath,amssymb}
\usepackage[caption=false]{subfig}
\usepackage{hyperref,breakurl}
\begin{document}

\title{Unusual face of radiation friction: enhancing production of longitudinal plasma waves}
\author{E.G.~Gelfer}\affiliation{ELI Beamlines, Institute of Physics of the ASCR, v.v.i., Czech Republic}\affiliation{National Research Nuclear University ``MEPhI'' (Moscow Engineering Physics Institute), 115409 Moscow, Russia}
\author{N.V.~Elkina}\affiliation{Helmholtz-Institut Jena, 07743 Jena, Germany }
\author{A.M.~Fedotov}\affiliation{National Research Nuclear University ``MEPhI'' (Moscow Engineering Physics Institute), 115409 Moscow, Russia}
\begin{abstract}
We study the penetration of ultraintense circularly polarized laser pulses into a thick subcritical plasma layer with accounting for radiation friction. We show that radiation pressure is enhanced by
radiation friction in the direction transverse to the laser pulse propagation, and that for stronger and longer laser pulses this mechanism dominates over the ordinary ponderomotive pressure, thus resulting in a stronger charge separation than anticipated previously. We give estimates of the effect and compare them with the results of 1D and 2D PIC simulations. 
\end{abstract}
\pacs{41.60.Ap, 52.38.Kd, 41.75.Jv}
\keywords{radiation friction, radiation pressure, laser plasma, charge separation, ion acceleration}
\maketitle

A new generation of 10 PW laser facilities (e.g., ELI Beamlines \cite{ELI}, Apollon \cite{Apollon}, ELI NP \cite{ELINP}) will be soon commissioned around the world, providing very strong fields with dimensionless amplitude $a_0=\frac{eE}{m\omega c}$ of the order of several hundreds. Here $-e$ and $m$ are electron charge and mass, $\omega$ is the laser carrier frequency, $E$ is the electric field amplitude, and $c$ is the speed of light. For $a_0\gg1$ the electron quiver motion is already ultrarelativistic, but as $a_0$ approaches few hundreds, it should become also strongly affected by radiation friction (RF) \cite{RDR2004}. 

For this reason, RF impacts on various laser plasma interaction processes and dynamics (nonlinear Thomson and Compton scattering \cite{compton}, inverse Faraday effect \cite{liseykina2016}, transform of electron bunches crossing a laser pulse \cite{beams_cross_pulse}, radiative trapping of electrons \cite{rad_trapping,elkina2014}, etc.) have received recently a substantial attention, see the reviews \cite{reviews}. Analytical solutions for a single particle motion with RF included are known \cite{RFpapers,zeldovich1975,bulanov2011,Yaremko,elkina2014} for such simple cases as a constant magnetic field, a uniformly rotating electric field, and a plane wave field, however most of the research was performed using numerical simulations (different numerical approaches are compared in \cite{vranic2016}). One of the most promising applications of powerful lasers is ion acceleration in a plasma. The ions are accelerated by a quasistatic electric field arising because of the charge separation created by the laser pulse, for a review and recent experimental results see Refs.~\cite{ion_acc_rev,ion_acc_exp}. Two theoretical models are discussed -- the light sail regime \cite{bulanov2010,light_sail} for thin targets and the hole boring regime \cite{hole_boring} for thick ones.

In this Letter we study the \textit{impact of RF} on longitudinal field generation \cite{bulanov1989} by circularly polarized (CP) gaussian laser pulses propagating in a thick cold plasma with \textit{immobile ions}. To facilitate the penetration of electrons inside the high field region experiencing fully the action of RF force, we consider much lower plasma densities than in the previous studies \cite{zhidkov2002,tamburini2010,tamburini2011,brady2012,nakamura2012,bashinov2013,stark2016,micha2012
}. For simulations we modified the PIC code EPOCH \cite{EPOCH} by including the classical RF into the particle pusher as described in detail in Refs.~\cite{zhidkov2002,tamburini2010}. 

\begin{figure*}[th!]
\subfloat{\includegraphics[width = 0.5\linewidth]{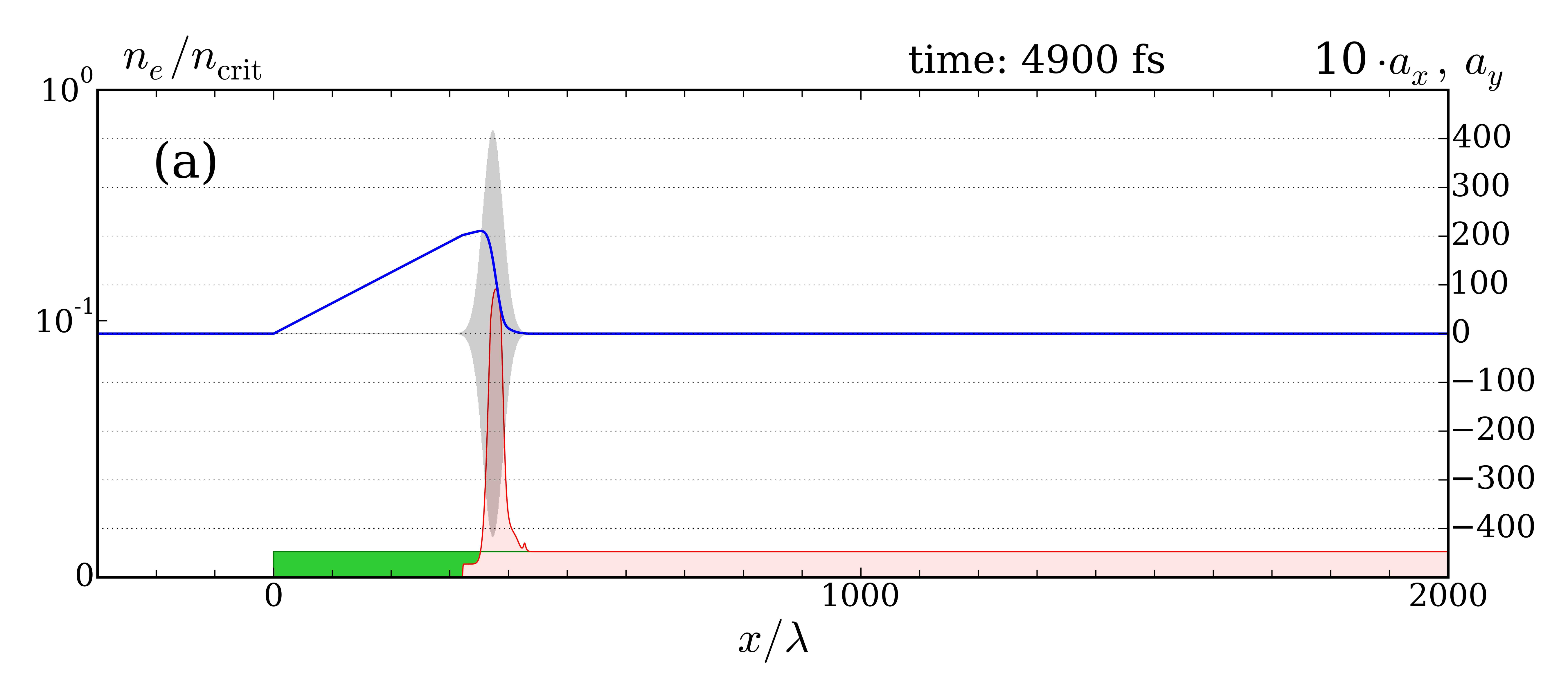}} 
\subfloat{\includegraphics[width = 0.5\linewidth]{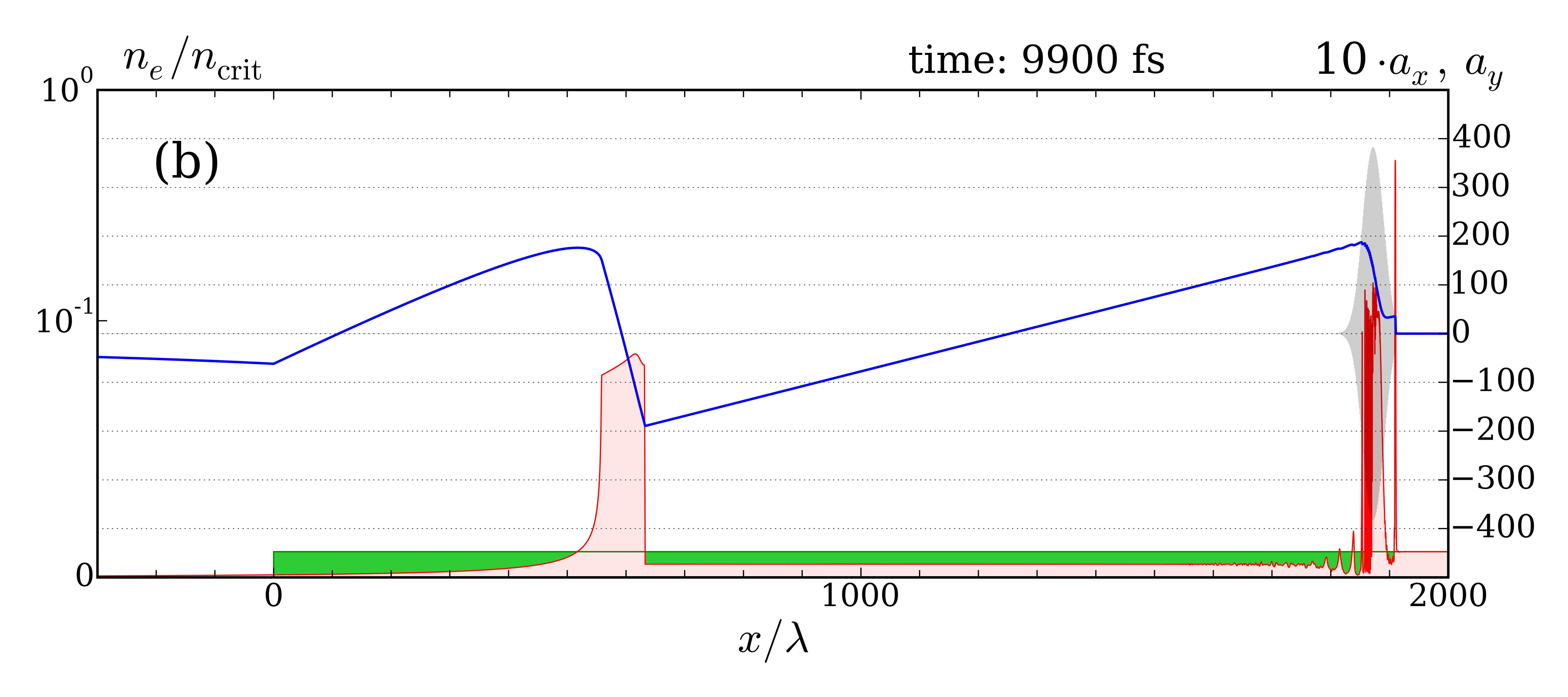}}\\ 
\subfloat{\includegraphics[width = 0.5\linewidth]{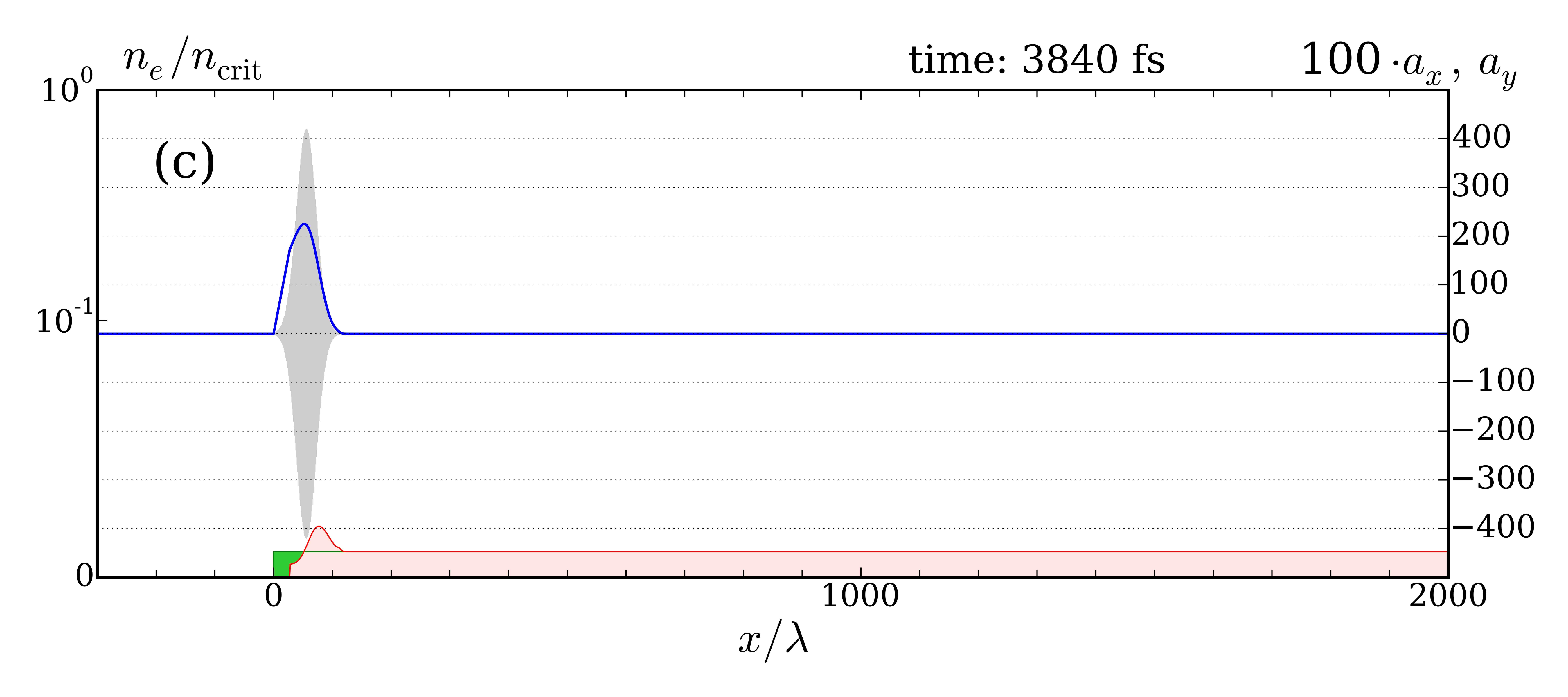}} 
\subfloat{\includegraphics[width = 0.5\linewidth]{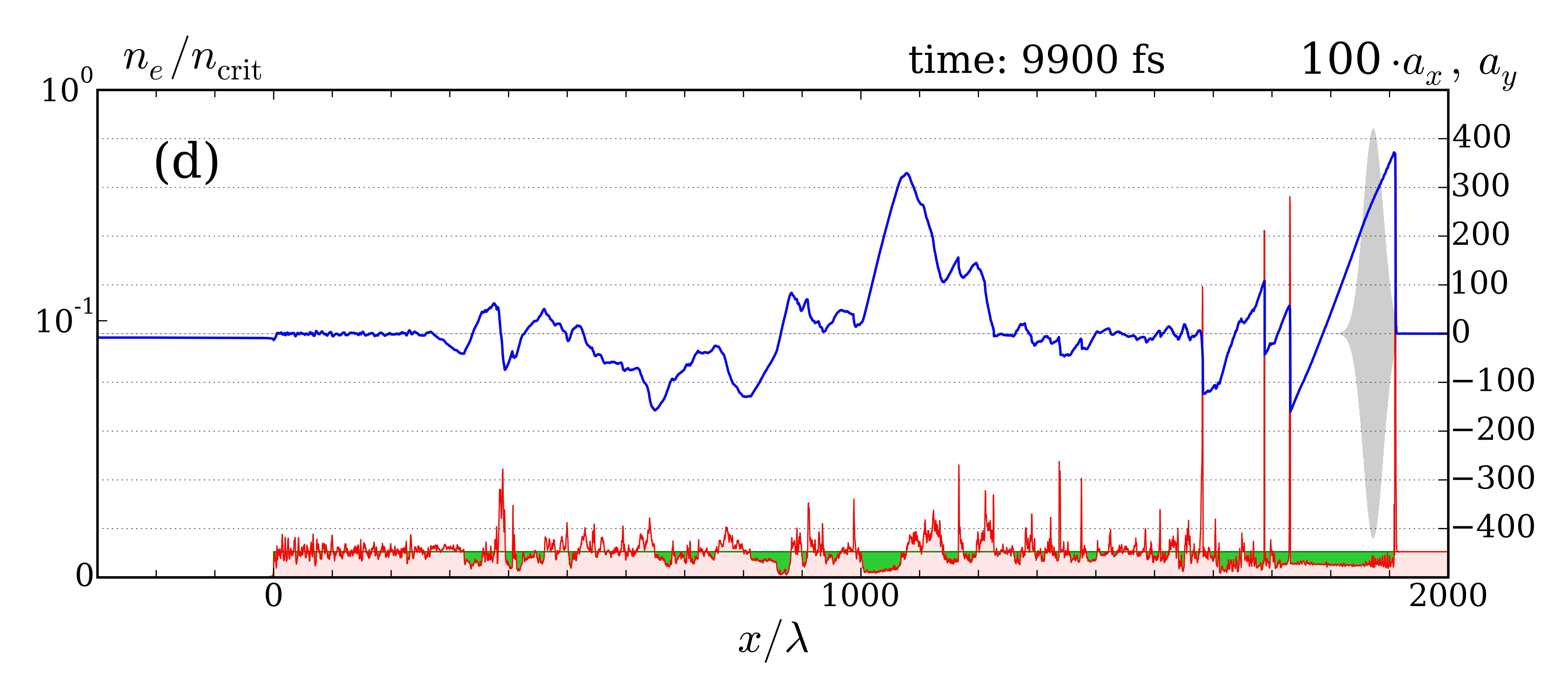}}
\caption{\label{fig1}
(Color online). Successive snapshots of a circularly polarized laser pulse propagating through a 1D plasma with [(a)--(b)] and without [(c)--(d)] RF: red and green filled areas -- electron and ion densities in units of critical density; grey line -- $y$ component of the dimensionless transverse electric field; blue curve -- longitudinal component of the dimensionless electric field multiplied by factors of $10$ for (a)--(b) and $100$ for (c)--(d). Laser and plasma parameters are given in the text on p.~\pageref{params}.}
\end{figure*}

To figure out the role of RF, let us start by presenting in Fig.~\ref{fig1} the results of 1D simulations of a laser pulse propagation in a cold plasma \textit{with} and \textit{without} RF. The values of the parameters are picked up according to the expectations of upcoming attainability, e.g., at ELI Beamlines \cite{ELI}: peak envelope amplitude $a_0=420$ (corresponding to peak intensity $I_{\text{L}}=5\cdot 10^{23}$ W/cm$^2$), full duration half maximum (FDHM) $t_{\text{p}}=125$fs, and wavelength $\lambda=1\mu$m. The electron density of the undisturbed plasma with ion charge number $Z=1$ is $n=0.01n_c$, where $n_c=m\omega^2/4\pi e^2$ is the critical density. The 1D simulations were performed with $100$ cells/$\lambda$ and $20$ particles per cell.\label{params}

As the laser pulse enters the plasma, it grabs all the electrons on its front, no matter whether RF is taken into account or not. The resulting charge separation creates a quasistatic longitudinal electric field of strength $E_\parallel(x)=4\pi enx$ for $0<x<x_{\text{m}}$, where $x_{\text{m}}$ is the leftmost position of the shifted electrons. This field pulls back the electrons and its amplitude is growing as the pulse penetrates deeper into the plasma until a breakdown at $t=t_{\text{bd}}$, when a bunch of electrons eventually penetrates back through the pulse, starting to accelerate against [see Fig.~\ref{fig1}~(a) and~(c)]. Such bunches, generated at the successive breakdowns and running leftwards, partially screen the electrostatic field, thus bounding its amplitude, see Fig.~\ref{fig1} (b) and (d). Comparing the two cases: when RF is turned on [Fig.~\ref{fig1}~(a)--(b)], and off [Fig.~\ref{fig1}~(c)--(d), let us emphasize that in this case the scale of the longitudinal field is additionally stretched by a factor of ten], one observes that the amplitude and period of a longitudinal plasma wave are both much higher when RF is taken into account. 

To explain the apparent paradox: radiation friction \textit{enhances} longitudinal acceleration of the electrons (see also \cite{RFpapers,zeldovich1975}), we propose a model for the initial stage of the process before the first breakdown. Since $n\ll n_c$, let us consider motion of a single leftmost electron driven by the transverse field of the pulse
\begin{equation}
\boldsymbol{a}_\bot=a_0(\varphi)\{0,\cos\varphi,\sin\varphi\},\quad \varphi=\omega(t-x/c),
\end{equation}
and the longitudinal field $a_\parallel=-\tilde{n}\xi$ of the naked ions, where  $\xi=\omega x_{\text{m}}/c$ and $\tilde{n}=n/n_c$. Equations of motion in these dimensionless variables read
\begin{equation}\label{eqm}
\begin{split}
\frac{d\mathbf{u}_\bot}{d\tau}=(1-\beta_x)\boldsymbol{a}_\bot-\mu a_0^2\gamma^2(1-\beta_x)^2\boldsymbol{\beta}_\bot,\\
\frac{du_x}{d\tau}=\boldsymbol{a}_\bot\boldsymbol{\beta}_\bot-\mu a_0^2\gamma^2(1-\beta_x)^2\beta_x-\tilde{n}\xi.
\end{split}
\end{equation}
Here $\mathbf{u}=\gamma\boldsymbol{\beta}$ is the electron 4-velocity spatial component, $\boldsymbol{\beta}=\mathbf{v}/c$,  $\tau=\omega t$, $\gamma=(1+u_x^2+u_\perp^2)^{1/2}\equiv[(1+u_\perp^2)/(1-\beta_x^2)]^{1/2}$, $\mu=2\omega r_e/3c\simeq 1.18\cdot 10^{-8}$, $r_e=e^2/mc^2$ is the classical electron radius, and we retain only the dominant ($\propto\gamma^2$) contribution to RF force in the Landau-Lifshitz form \cite{LL}.

Using $d\varphi=d\tau(1-\beta_x)$ and assuming that radiation damping is weak ($u_\bot\approx a_0\gg 1$), we express the transverse field $\boldsymbol{a}_\bot$ from the first of Eqs. (\ref{eqm}) and substitute it into the second one, thus arriving at 
\begin{equation}\label{eq1}
\frac{du_x}{d\tau}=\frac{1}{2\gamma}\frac{d a_0^2}{d\varphi}+\mu a_0^4\frac{1-\beta_x}{1+\beta_x}-\xi\tilde{n}.
\end{equation}
Here the first two terms on the RHS jointly describe the effect of radiation pressure [compare to Eq.~(9) of Ref.~\cite{bashinov2013}]. The first of them is the conventional relativistic \textit{ponderomotive force} \cite{bauer1995}, while the second one is \textit{induced by RF} as follows: RF modifies the transverse quiver motion by additionally increasing the angle between the electron momentum and the magnetic field, thus enhancing the longitudinal component of the accelerating Lorentz force $\propto\mathbf{v}\times\mathbf{B}$ \cite{zeldovich1975}. The second term is precisely the resultant of this Lorentz force increment (directed forward) and the longitudinal RF force (directed backward). Alternatively, it can be understood as the net gain of momentum flux in a Thomson scattering occurring because the momenta of all the absorbed photons are parallel to $x$ axis, while the scattered photons are bended by angles $\simeq \gamma^{-1}$ \cite{LL}. The unusual stronger scaling $\propto a_0^4$ is due to the transverse electron motion \cite{LL,bul2015}. To avoid possible confusion, let us stress that the RF-induced accelerating force [the second term on the RHS of Eq.(\ref{eq1})], though at first glance might seem reminiscent to the radiation pressure force $\propto a_0^2$ proposed in Ref.~\cite{bulanov2010} to describe unlimited ion acceleration in the light sail regime, is in fact completely different, as is derivable from the Landau-Lifshitz equation for a \textit{dilute} plasma without any account for plasma effects. In contrast, the radiation pressure considered in Ref.~\cite{bulanov2010} originated as a combination of what we here call ponderomotive force, and the purely plasma effect of laser pulse reflection from the \textit{opaque} plasma layer. 

Though Eq.~(\ref{eq1}) does not admit an exact solution, the process clearly splits into \textit{stages}, thus allowing us to carry out a qualitative analysis and propose some estimates. Initially, as the pulse just starts penetrating into a plasma, the charge separation $\xi$ is small, the Coulomb force is negligible, and the electrons are \textit{accelerated} by radiation pressure. However, after some time $t_{\text{acc}}$, when the Coulomb force counterbalances the radiation pressure, the process enters the stage of \textit{steady deceleration} and the LHS of Eq.~(\ref{eq1}) can be neglected. This stage lasts until the \textit{breakdown}, when a bunch of electrons finally penetrates to the rear of the  pulse. We assume that longitudinal motion of the leftmost electrons is ultrarelativistic [$u_x\gg u_\perp$, $\xi(\tau)\approx\tau$], in such a case $t_{\text{acc}}\lesssim t_{\text{bd}}$, hence $t_{\text{bd}}$ estimates the period of the resulting longitudinal plasma wave. The time of breakdown $\tau_{\text{bd}}=\omega t_{\text{bd}}$ is fixed by
\begin{equation}\label{tbd}
T=\varphi(\tau_{\text{bd}})= \int_0^{\tau_{\text{bd}}}(1-\beta_x)d\tau\approx  \frac{a_0^2}{2}\int_0^{\tau_{\text{bd}}}\frac{d\tau}{u_x^2},
\end{equation}
where $T=\omega t_{\text{p}}$ is the dimensionless pulse duration.

Let us make the estimates, assuming in turn that one of the two competing mechanisms of radiation pressure (ponderomotive vs. RF-induced) dominates over the other. When the ponderomotive mechanism (PM) is dominant, the second term on the RHS of Eq.~(\ref{eq1}) can be neglected. Then, with a suggestive estimate $da_0^2/d\varphi\sim a_0^2/T$ in Eq.~(\ref{eq1}), we have $u_x(\tau)\simeq a_0^2/T\tilde{n}\tau$ for the steady deceleration stage $\tau_{\text{acc}}\lesssim \tau < \tau_{\text{bd}}$ and, substituting it further into Eq.~(\ref{tbd}), one finally obtains:
\begin{equation}\label{taccp}
\tau_{\text{bd}}^{\text{(PM)}}\simeq\left(\frac{a_0}{\tilde{n}\sqrt{T}}\right)^{2/3},\quad a_\parallel^{\text{(PM)}}\simeq \tilde{n}\tau_{\text{bd}}^{\text{(PM)}}\simeq a_0^{2/3}\left(\frac{\tilde{n}}{T}\right)^{1/3}
\end{equation}

In the opposite case of the RF-induced mechanism (RFM) we drop the first term on the RHS in Eq.~(\ref{eq1}) and complete the rest of the estimates following the same lines. In particular, for the deceleration stage we have $u_x(\tau)\simeq a_0^3\sqrt{\mu/\tilde{n}\tau}$, and furthermore
\begin{equation}\label{taccrr}
\tau_{\text{bd}}^{\text{(RFM)}}\simeq \sqrt{\frac{\mu T}{\tilde{n}}}a_0^2,\quad a_\parallel^{\text{(RFM)}}\simeq\sqrt{\mu\tilde{n}T}a_0^2.
\end{equation}

Equations (\ref{taccp}) and (\ref{taccrr}) estimate the wavelength and the amplitude of the resulting longitudinal wave in the PM and RFM dominated regimes, respectively. They can be used, in particular, to conclude that RFM outperforms PM ($a_\parallel^{\text{(RFM)}}\gtrsim a_\parallel^{\text{(PM)}}$) if 
\begin{equation}\label{c1}
\mu^3\tilde{n}T^5a_0^8\gtrsim1,
\end{equation}
i.e., for denser plasma and stronger and longer pulses. 

Let us briefly comment on the restrictions validating the assumptions of our derivation. The first one ($u_\perp\simeq a_0$), ensuring the weakness of the transverse motion damping due to RF, is that the transverse Lorentz force should substantially exceed the RF force. It turns out that this criterion can be formulated equivalently by that the energy stored in the resulting quasistatic longitudinal field remains much smaller than the total energy of the pulse, $\left(a_\parallel^{\text{(RFM)}}\right)^2\tau_{\text{bd}}^{\text{(RFM)}}\ll a_0^2T$. This implies that the energy of the accelerated electron bunch also remains always smaller, meaning that the conversion of a transverse alternating field into longitudinal quasistatic field is rather efficient. The second restriction $u_x(\tau_{\text{bd}})\gtrsim a_0$ is needed to ensure that longitudinal electron motion is relativistic. Using Eqs.~(\ref{taccrr}), these restrictions can be formulated explicitly as
\begin{equation}\label{c2}
\mu^3\tilde{n}T a_0^8\ll 1,\quad \tilde{n}T/\mu a_0^4\lesssim 1.
\end{equation}
For $a_0\gtrsim \mu^{-1/3}\approx440$ the first among the conditions (\ref{c2}) is the strongest and while $T\gg 1$ it also does not contradict Eq.~(\ref{c1}). For example, for $T\simeq 10^2$ and $100<a_0<500$ the restrictions (\ref{c1}) and (\ref{c2}) are fulfilled for $10^{-2}\lesssim \tilde{n}\lesssim 1$, which explains our choice of the simulation parameters for this Letter.

\begin{figure}[th!]
\includegraphics[width =\linewidth]{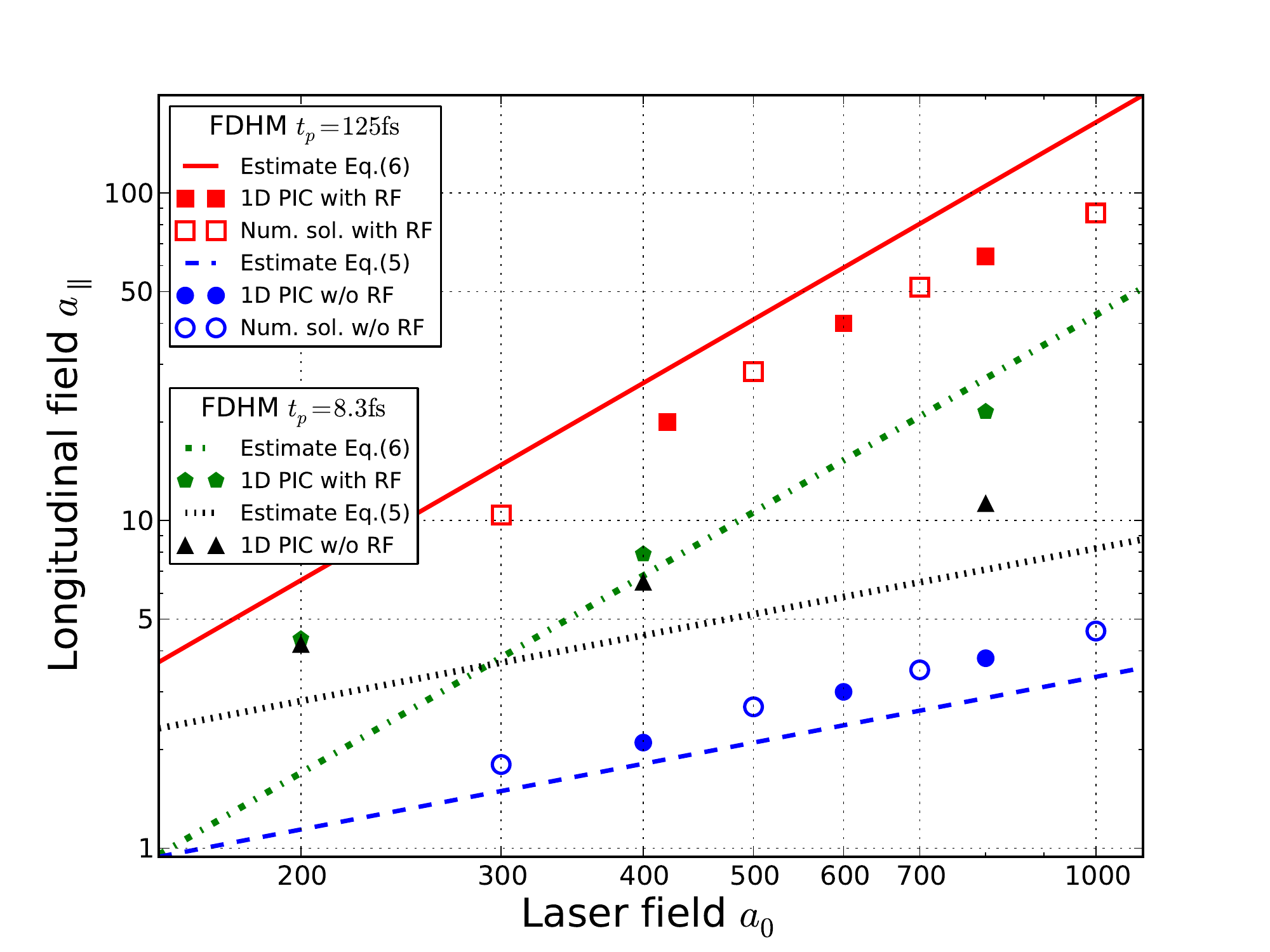}
\caption{\label{fig2}
(Color online). Amplitude $a_\parallel$ of a longitudinal wave generated by a CP laser pulse in a plasma with $Z=1$ and $n=0.01 n_c$ vs the amplitude $a_0$ of the driving laser pulse.}
\end{figure} 

\begin{figure*}[th!]
\subfloat{\includegraphics[width = 0.5\linewidth]{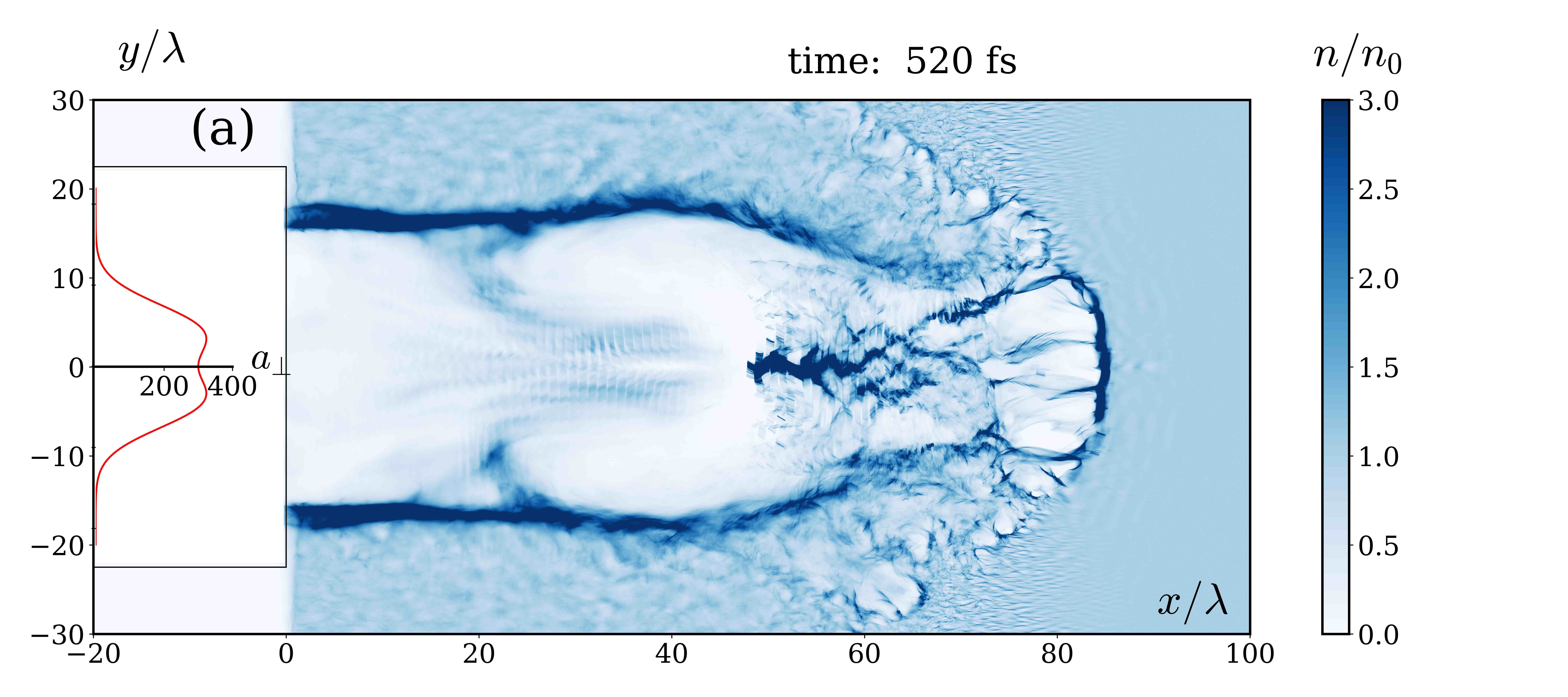}} 
\subfloat{\includegraphics[width = 0.5\linewidth]{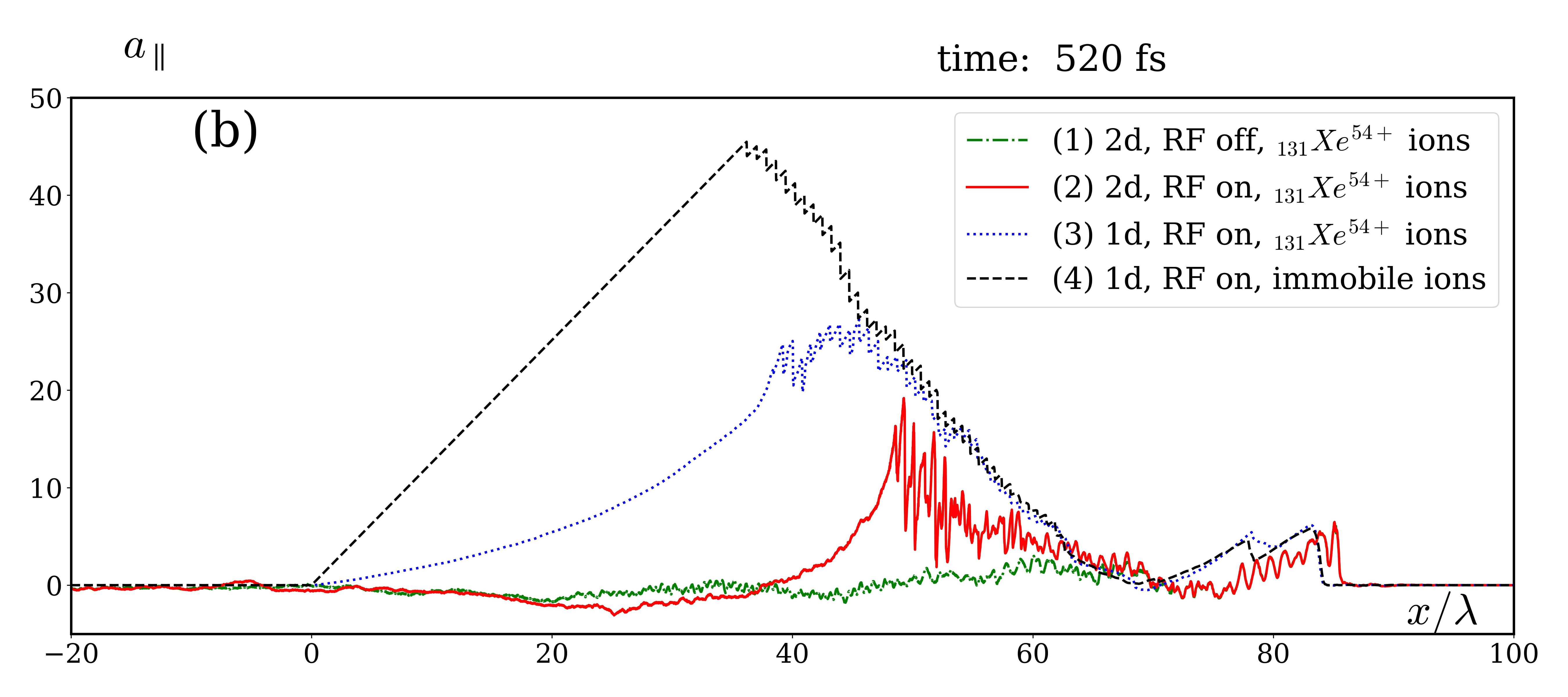}} \\
\subfloat{\includegraphics[width = 0.5\linewidth]{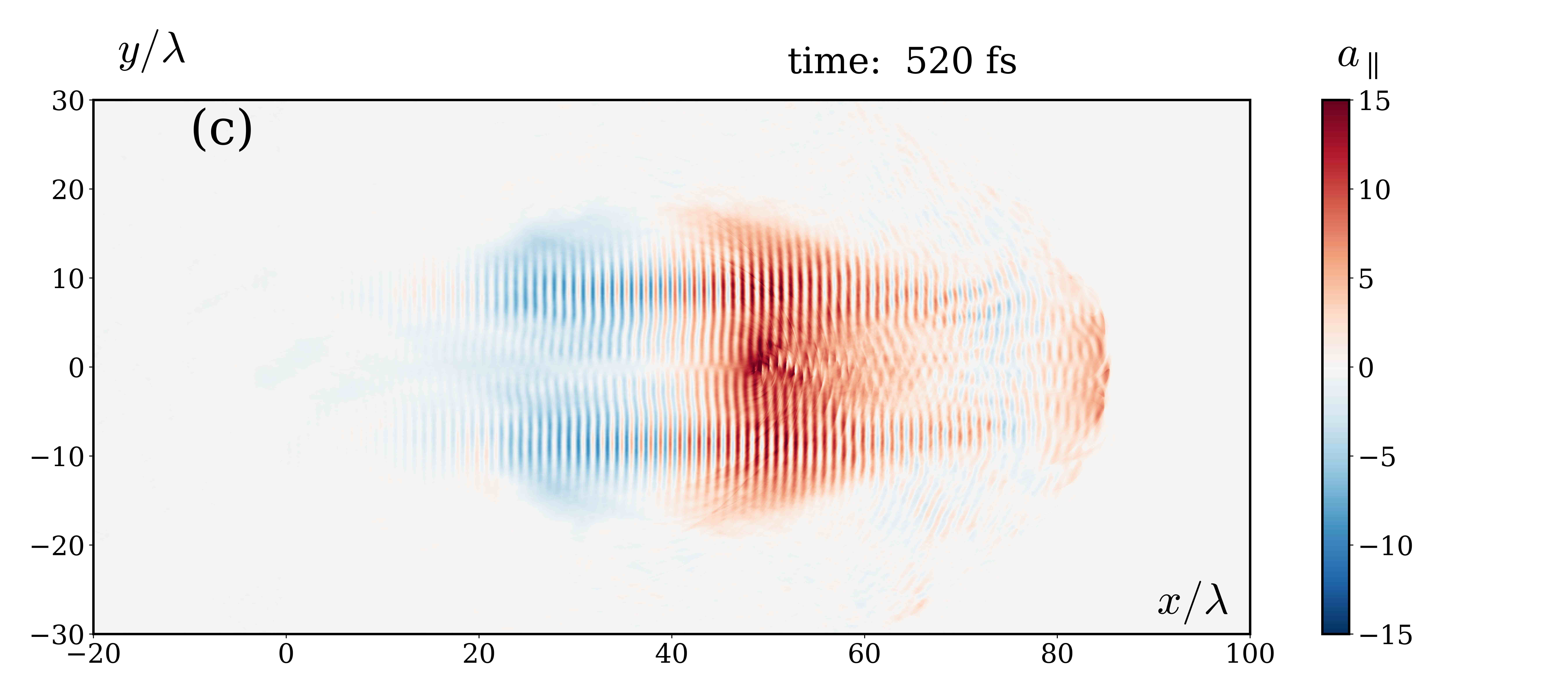}} 
\subfloat{\includegraphics[width = 0.5\linewidth]{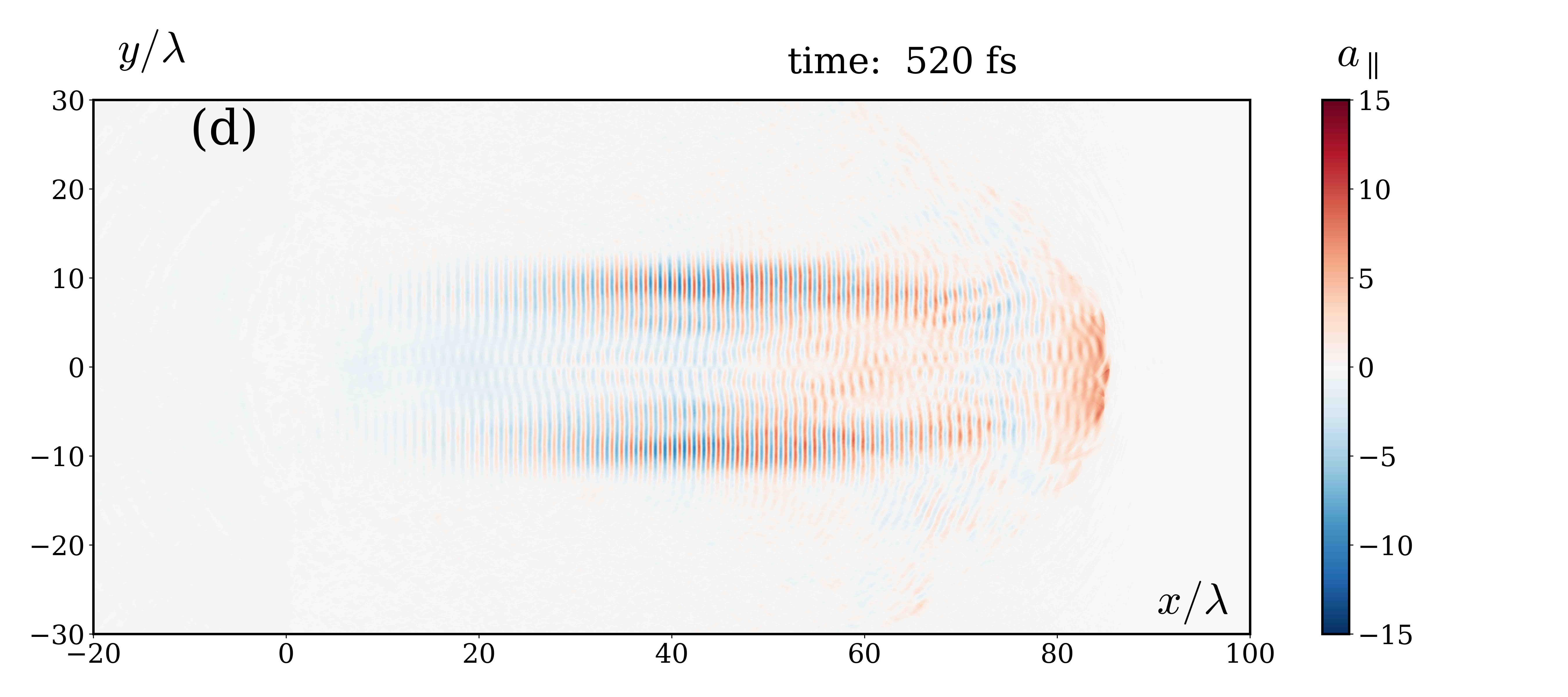}}  
\caption{\label{fig3}
(Color online). 2D simulations of propagation of a CP symmetric bimodal gaussian pulse in a plasma with mobile ions [parameters are given in the text on p.~\pageref{param2D}]: (a) -- electron density (inset: transverse pulse profile); (b) -- longitudinal field on the $x$-axis with (red solid line 2) and without RF (green dash-dot line 1), compared to 1D simulations with mobile (blue dotted line 3) and immobile (grey dashed line 4) ions, and with the same amplitude of the driving laser field on the $x$-axis; (c)  -- longitudinal field with RF; (d) -- longitudinal field without RF.}
\end{figure*}
 
In Figure \ref{fig2} we compare our estimates (\ref{taccp}) and (\ref{taccrr}) (which with the adopted double-log scale appear as straight lines) to the numerical solution of the original Eqs.~(\ref{eqm}) (uncolored bullets), as well as to the results of 1D PIC simulation (filled bullets), each performed with and without RF -- for short (FDHW$=8.3$fs) and long (FDHW$=125$fs) pulses. Corresponding data are shown in the same color. First of all, the figure shows that the markers for the numerical solution of our model (\ref{eqm}) are in good agreement with PIC simulations. This was expected, as the plasma effects should be negligible for the low densities considered thus far. Less trivially, as a rule the slopes of the curves and of the respective scatter data coincide, thus validating our estimates (\ref{taccp}) and (\ref{taccrr}) up to numerical coefficients $\sim1$. The only exceptions are the right upper square at the border of the applicability region (\ref{c2}), and the leftmost two pentagons shifted upwards from below the green dot-dashed curve due to transition to the PM-dominated regime to the left of its crossing with the black dot curve. Apart from that weak field region for a shorter pulse, and in the whole range of $a_0$ for longer pulses, RFM clearly dominates over the PM. The established correspondence between the three approaches confirms both our estimates and the accuracy of our numerics.

To substantiate the effect in a more realistic setup, we also made 2D EPOCH simulations with mobile ions ${}_{131}^{54}\text{Xe}^{54+}$, which we assume fully ionized according to the rough estimates based on Ref.~\cite{poprz}. The 2D simulations were performed with $100$ cells/$\lambda$ and $10$ particles per cell. The results for a laser pulse \label{param2D} of intensity $a_0=300$ ($I_L\simeq 2.5\cdot 10^{23}$W/cm$^2$), FDHM $t_{\text{p}}=125$fs and waist radius $w=5\lambda$,  focused at the left plasma boundary, are summarized in Figs.~\ref{fig3} (a)--(d). To prevent immediate transverse expel of electrons from the pulse front we used pulses with symmetric bimodal gaussian transverse profile shown in the inset of Fig.~\ref{fig3}~(a), with such distance between the peaks that the maximal field at the $x$-axis coincides to the peak envelop amplitude of each superposed pulses. Also, we increased the plasma density $n$ to $0.2n_c$ in order to strengthen the quasistatic longitudinal field on a background of the alternating longitudinal field of the pulse attributed to its tight focusing. The longitudinal fields computed with and without RF are compared in Figs.~\ref{fig3}~(c) and (d), where one can observe that the effect is extremely well pronounced in 2D. Moreover, the longitudinal field distribution on the $x$-axis is also in a qualitative agreement with 1D simulations, see Fig.~\ref{fig3}~(b). The most notable 2D effect is that part of the electrons bypasses the ion bubble \cite{bubbles}, getting inside from its rear side [see Fig.~\ref{fig3}~(a)], and in this way screening the quasistatic longitudinal field. Its decrease (as compared to the 1D simulation) at the rear of the resulting longitudinal wave in Fig.~\ref{fig3}~(b) is explained in part by this effect (compare the lines 2 and 3), and in the rest part by decrease of the charge separation gap due to expel of the ions (compare the lines 3 and 4).

To conclude, we propose a new mechanism of quasistatic longitudinal plasma field generation by laser pulses. The mechanism is based on \textit{enhancing} the longitudinal Lorenz force by transverse \textit{radiation friction}, and for long and intense pulses considerably outperforms the conventional ponderomotive pressure. Though less pronounced, the effect remains feasible for the parameters of the upcoming ELI Beamlines facility. Further development of our model by taking into account ion mobility and its application as a novel alternative mechanism for ion acceleration, will be given in a separate forthcoming publication.

EGG and AMF are grateful to S.V. Bulanov, S.S. Bulanov, S. Rykovanov, F. Mackenroth, M. Grech, T. Esirkepov, S. Bochkarev, E. Nerush, S. Popruzhenko, M. Vranic, G. Korn, O. Klimo, and S. Weber for valuable discussions, and to K. Krylov and E. Echkina, in addition, for advising on software and technical assistance. The research was performed using the code EPOCH (developed under the UK EPSRC grants \texttt{\detokenize{EP/G054940/1}}, \texttt{\detokenize{EP/G055165/1}}, and \texttt{\detokenize{EP/G056803/1}}) and resources of the NRNU MEPhI high-performance computing center, and was partially supported by the MEPhI Academic Excellence Project (Contract No.~\texttt{\detokenize{02.a03.21.0005}}), the Russian Fund for Basic Research (Grants~\texttt{\detokenize{16-32-00863mol_a}} and \texttt{\detokenize{16-02-00963a}}), the project ELITAS (ELI Tools for Advanced Simulation) \texttt{\detokenize{CZ.02.1.01/0.0/0.0/16_013/0001793}} from European Regional Development Fund, and the Helmholtz Association (Helmholtz Young Investigators group VH-NG-1037). The preliminary 2D simulations (not presented in the paper) were also verified with another open source code SMILEI \cite{smilei} (courtesy of M. Grech).

\end{document}